\documentclass[preprint,12pt]{elsarticle}

\usepackage{amssymb}
\usepackage{subcaption}
\usepackage{times}
\usepackage{epsfig}
\usepackage{graphicx}
\usepackage{amsmath}
\usepackage{amssymb}
\usepackage{float}
\usepackage{makecell}
\usepackage{color}
\usepackage{multirow}
\usepackage{graphicx,placeins}
\usepackage[T1]{fontenc}
\newcommand{\latinphrase}[1]{\textit{#1}} 
\newcommand{\etal}{\latinphrase{et~al.}\xspace}
\newcommand{\ie}{\latinphrase{i.e.,}\xspace}

\newcommand{\eg}{\latinphrase{e.g.}\xspace}

\usepackage{times}
\usepackage{epsfig}
\usepackage{graphicx,subcaption}
\usepackage{amsmath}
\usepackage{amssymb}
\usepackage{lipsum}
\usepackage{mathtools, nccmath}

\usepackage{amsmath}

\usepackage[linesnumbered,ruled,vlined]{algorithm2e}
\PassOptionsToPackage{hyphens}{url}\usepackage{hyperref}

\usepackage[cmintegrals]{newtxmath}
\makeatother

\begin{document}

\begin{frontmatter}



\title{PLLM-CS: Pre-trained Large Language Model (LLM) for Cyber Threat Detection in Satellite Networks}


\author[inst1]{Mohammed~Hassanin}
\author[inst2]{Marwa Keshk}
\author[inst2]{Sara Salim}
\author[inst3]{Majid Alsubaie}
\author[inst3]{Dharmendra Sharma}

\affiliation[inst1]{organization={the University of South Australia (UniSA)},
            state={SA},
            country={Australia}}

\affiliation[inst2]{organization={University of New South Wales},
            state={Canberra},
            country={Australia}}
\affiliation[inst3]{organization={University of Canberra},
            state={Canberra},
            country={Australia}}

\begin{abstract}

Satellite networks are vital in facilitating communication services for various critical infrastructures. These networks can seamlessly integrate with a diverse array of systems. However, some of these systems are vulnerable due to the absence of effective intrusion detection systems, which can be attributed to limited research and the high costs associated with deploying, fine-tuning, monitoring, and responding to security breaches. To address these challenges, we propose a pre-trained Large Language Model for Cyber Security , for short PLLM-CS, which is a variant of pre-trained Transformers \cite{transformers}, which includes a specialized module for transforming network data into contextually suitable inputs. This transformation enables the proposed LLM to encode contextual information within the cyber data.
To validate the efficacy of the proposed method, we conducted empirical experiments using two publicly available network datasets, UNSW\_NB 15 and TON\_IoT, both providing Internet of Things (IoT)-based traffic data.  Our experiments demonstrate that proposed LLM method outperforms state-of-the-art techniques such as BiLSTM, GRU, and CNN. Notably, the PLLM-CS method achieves an outstanding accuracy level of 100\% on the UNSW\_NB 15 dataset, setting a new standard for benchmark performance in this domain.


\end{abstract}




\end{frontmatter}


\section{Introduction}
Recent advances in satellite communications have progressed the development of many end-user services, including the Internet of Things (IoT), Internet of Vehicles (IoV) and healthcare. These systems provide opportunities to develop physical applications such as smart cities and enterprise management systems. Satellites are complicated devices that provide many services and tasks. Satellites offer a means to extend wireless networks to unreachable places to terrestrial infrastructures \cite{de2015satellite}. One method of classifying satellites is the distance of their orbit from Earth relative to each other. These range from low earth orbit (LEO), which is closest to Earth, to geostationary earth orbit (GEO), which is the furthest from Earth. GEO orbits are slower with a wide orbital path, whilst LEO orbits are faster.

Satellites are launched for different purposes and missions. They are platforms for performing tasks based on their in-built equipment and sensors. Some are used to monitor and send images of the Earth to detect environmental changes. Others provide internet services to remote areas and facilitate applications, including healthcare emergencies and self-driving cars  \cite{badue2021self}. However, all have shared entities for providing basic services; for instance, data processing information collected from their sensors is an initial step for them. Their processes detect their orientations and positions, malfunctions, and diagnoses. Some actuators responsible for charging satellites using solar radiation have to be equipped with panels in their systems \cite{colagrossi2022spacecraft}.

Satellite networks are vulnerable to cyberattacks, like many other systems, such as IoT and IoV. These threats become more severe when networks’ data are dependent on physical devices such as satellites which, in turn, require more robust Intrusion Detection Systems (IDS); for example, according to the study in \cite{iot_risks}, 57\% of IoT devices are exposed to severe attacks. Attacking satellites is more dangerous than attacking the IoT and other networks because they are vital for remote areas such as army units. Moreover, the cutting of connections among distant military units by satellite attacks can cause breakdowns in command and control, as demonstrated in the recent conflict between Ukraine and Russia  \cite{satellite_risks}. If satellite networks are integrated with IoT devices, another gate is open for cyberattacks; for instance, a Mirai attack  \cite{kolias2017ddos} based on a botnet along with a Distributed Denial of Service (DDoS) can exploit communications, such as data transport systems, to cause cyber threats.

Satellite networks are composed mainly of ground stations, space segments and up-and-down links operated from the ground segments. The following attacks/adversaries can confront them. 1) a DoS based on overwhelming the target with dense traffic to prevent legitimate users from normal access  \cite{wankhede2018attack}. It causes inaccessibility to resources and/or, ultimately, failure of a service. As reported in  \cite{verisign}, there is steady growth in the number of DoS attacks, sizes, frequencies and complexities. A DoS has various types, such as ICMP, UDP, SYN and HTTP flood. Although different approaches use these types, all result in the final inaccessibility of the targeted system.

A DoS is a multidisciplinary attack that can be used in all network connections, including satellite ones. 2) A Distributed DDoS attack harmonizes multiple DoS ones simultaneously to invade a single target system., It has more resistance and complexity than a DoS one due to its multiple instances with a single aim. As a result, it is more severe and disruptive because it can access more resources, and it is unrealistic to shut down all infected systems simultaneously. Such attacks are related to terrestrial networks, as detailed in \cite{lau2000distributed, wood2002denial, borisov2007denial}. However, they are still disruptive to SSNs because LSN networks can be attacked globally, and SSNs provide global information. Also, they have other features, including low latency, limited numbers of clients and sparse connectivity between ground and space segments which make an attacker’s task easy.

To address the above-mentioned issues, Smart Satellite Networks (SSN) that integrate
satellite systems, IoT over network communications, and Machine Learning paradigms have been proposed \cite{li2020distributed, moustafa2022dfsat}. One direction, especially for military space systems, is encryption-based methods  \cite{jackson2018exploring, o2016secure}. In \cite{ostad2019efficient}, elliptic curve cryptography is proposed for securing satellite communications. However, these methods secure only the physical layer and IDSs for SSNs have been investigated on a limited scale. In one study \cite{zhao2020intelligent},  a Convolutional Neural Network (CNN) and Long Short-term/Temporary Memory (LSTM) models are used inside a federated learning architecture to detect adversaries. 

In this paper, we propose using self-attention modules \cite{transformers} to build a robust system capable of detecting adversaries with high levels of accuracy. The main contribution is the development of a robust attention-based IDS, namely, a Pre-Trained Large Language Model for Cyber Security Defence (PLLM-CS), to determine the presence of advanced adversaries in SSNs. To our knowledge, this is the first study to propose a transformer-based method for detecting satellite adversaries. Although it is considered a centralized approach compared to a distributed one, it still outperforms the baselines. Also, it performs better than the benchmarks in detecting intrusions in any network data. It uses transformers at the model's core because of their capabilities to learn long-term contextual representations. Following are the contributions of this research.

\begin{itemize}
    \item Developing a Pre-Trained Large Language Model for Cyber Security Defence method as an IDS for SSNss to determine highly advanced adversaries, including fuzzer, DoS and reconnaissance attacks.
    \item Illustrating that this method is a generic solution for any network by validating it on satellite and IoT network data.
    \item Providing extensive experiments using various datasets with different attacks and comparing traditional and deep learning models.
\item Providing a new benchmark for IDSs on two publicly available datasets.

\end{itemize}

\section{Related Works}
\textbf{Satellite Network attacks}
Satellite Network Attacks: although satellite systems are similar to terrestrial ones  \textit{w.r.t} network data, they have different methodologies. As a DoS attack is popular in all types of networks, it is more serious when deployed in a satellite one because of its wide coverage and involvement of multiple technologies. However, a DoS adversary can target multiple things in SSNs, including the connections between their nodes, in the same way as attacking classical networks using millions of bots and botnets. In Coremelt  \cite{studer2009coremelt}, inter-domain links are targeted by adversaries to cause congestion. A number (N) of bots and a botnet are used to generate legitimately similar flows to bypass their adversaries. However, these adversaries can then initiate flows that bypass the links in the target, causing congestion in any link. Crossfire  \cite{kang2013crossfire} is similar to Coremelt \cite{studer2009coremelt} in its way of attacking. The links are congested and overloaded as the connections between a network’s topology are hindered. Coremelt and Crossfire are difficult to alleviate because they are indistinguishable from legitimate traffic. Another attack that targets a satellite’s up and down links is proposed in ICARUS  \cite{giuliari2021icarus}.  It generates legitimate traffic to overload the communications of an LSN and is considered the easiest because of its low bandwidth. It causes congestion to an ISL with more traffic and combines multiple links to construct a more complicated form. In general, it creates more threats than the previous two attacks because some of their mitigations are not practically applicable to it.

\textbf{Satellite IDS}
Satellite IDS: Recently, SSNs have progressed significantly in covering rural areas and providing cheap services for industrial sectors, such as healthcare and the IoT and IoV, particularly in the absence of wireless networks. However, they require robust IDSs to maintain the continuity of these services.  In \cite{na2018distributed}, Zhenyu \etal proposed a new mechanism based on the design of distributed routing LEO SSNs. It uses classical Machine Learning (ML) methods to estimate the network traffic and then intelligent routing decisions are made to maintain the traffic without any overloads.  Gunn \etal \cite{gunn2018anomaly} proposed using LSTM to detect any anomalies in network data which, as a result, reduces the number of false alarms in satellite communication systems. Another study in  \cite{pilastre2020anomaly} was proposed to monitor a spacecraft and maintain the health of its entire system. Dictionary learning and sparse representation detect intrusions in satellite network data.  Cheng \etal \cite{cheng2021research} used LSTM to predict anomalies in SSNs and then define a system's health as the difference between its actual and expected parameters.

In \cite{wang2022deep}, the authors proposed a method for detecting anomalies in satellite telemetry data. Firstly, the Deviation Divide Mean over Neighbors (DDMN) technique is used to detect intrusions in the data. Then, the LSTM  learns deep features from the multivariate data. Finally, a Gaussian model is employed to detect intrusions in the LSTM’s features. In the most recent study \cite{zeng2022satellite}, Zeng \etal proposed CN-FALSTM, a data-driven technique for detecting anomalies in the telemetry data of a satellite. Its main objective is to mitigate the false positive rates. Likewise, Yun \etal \cite{yun2022data} developed a model for predicting the voltage and current of a satellite in a low orbit. Most recently, Moustafa \etal \cite{moustafa2022dfsat} introduced a federated learning IDS based on LSTM to detect intrusions on satellite systems.

\textbf{Transformers as IDS}
Since Transformers\cite{transformers} have been proposed as an attention-based solution to different paradigms, such as vision and NLP, great progress has been made in all aspects of ML  \cite{hassanin2022crossformer, hassanin2022visual}.
In \cite{tan2019neural}, Tan \etal used an attention-based technique for real-time detection because of the time-slot capabilities of transformers. They compared bidirectional LSTM (BiLSTM) and Conditional Random Fields (CRFs) as baselines with their proposed method performing the best In a similar study  \cite{wu2022rtids}, Wu proposed using a typical transformer design consisting of positional encoding to identify tokens, encoders to learn low features and then self-attention modules to encode long-range relationships. In \cite{ghourabi2022security}, Ghourabi used different structures of transformers to detect intrusions in network data. It benefited from the power of self-attention modules to consider the context of the input data to protect healthcare systems from cyberattacks. In a recent study, Luo \etal \cite{luo2022hierarchical} introduced a modified fused architecture from a CNN and transformers to detect intrusions in network data which involved a proposed CNN-transformer NIDS and traffic spatiotemporal features. It uses softmax to encode the selection of soft features to improve the capability of the final model. Although these models are applied to different applications, such as network data and healthcare systems, no IDS for SSNs using the most recent technology of deep ML transformers has been investigated.



\section{Method}
\begin{figure*}[t!] 
\centering 
\includegraphics[width=\linewidth]{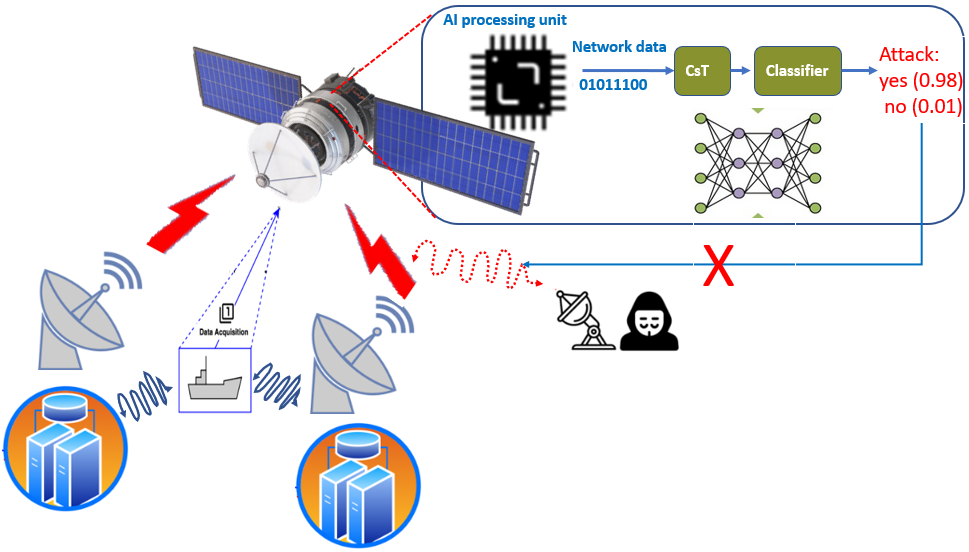}
   \caption{The visualization of the proposed method, PLLM-CS, as an IDS with the satellite network. PLLM-CS will be an embedded AI-processing unit such as Nvidia Jetson, plugged into the satellite.}
  \label{fig:architecture}
\end{figure*}

The architecture of the proposed method is shown in Figure \ref{fig:architecture}.

\subsection{Pre-Processing}
\begin{figure}[]
    \centering
    \includegraphics[width=\linewidth]{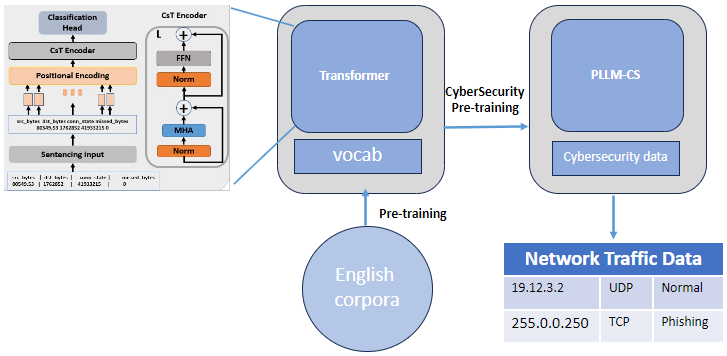}
   \caption{Visual explanation for PLLM-CS components. Firstly, the input is forwarded to the sentencing step to facilitate context encoding by MHA.}
  \label{fig:Cst}
\end{figure}

Initially, transformers aimed to help natural language-processing applications, such as text translations and summarizations  \cite{zaheer2020big}. In them, the text is a set of sentences, each of which is a set of words representing a token. However, cyber-security-network data are different because they are not sentences but multivariate series. This means that their contexts and long-range relationships are not used, limiting the power of transformers. To address this issue, we propose forming sentences from the multivariate series’ tokens and then dividing them into tokens to learn the long-term relationships among them. Put, as each input feature is considered a word, putting these features together is how to form a sentence (see Figure \ref{fig:Cst}).

\subsection{Pre-Trained Large Language Transformer Model}

Self-attention modules have improved the performance of ML tasks, including visual recognition \cite{hassanin2019new}, natural language processing, and multimodal ones. The most popular module of such self-attention architecture is Transformers \cite{transformers}. Despite its great success, less attention is paid to cyber-security paradigms. Following previous studies in the literature, \cite{transformers}, the input is first fed into a preprocessing step at the beginning and then goes through transformer encoders stacked on top of each other.

The input corresponds to the multivariate sequence $\{\textbf{x}_{i} \in \mathbb{R}^{J} | i = 1, ..., N\}$, where $J$ is the total number of variables in the sequence and $N$ is the total number. In the initial layer of the transformer, positional information is mixed with input patches as the identities of similar tokens, called patch embedding, and is calculated as:
\begin{equation}
\begin{split}
Z_0 &= [\textbf{x}_1 E; \textbf{x}_2 E; ..., \textbf{x}_P E] ,\\ 
&E \in \mathbb{R}^{(J) \times C},  Z_0 \in \mathbb{R}^{T \times C}
\end{split}
\end{equation}
where $C$ is the embedding dimension, and $T$ is the number of patches. 

The result of the previous step $Z_0$, patch embedding, is fed into the core step of Transformers, \ie the \textit{self-attention}. It implicitly learns the dependencies between the various tokens by encoding the relationships between the three main matrices $Q, K, V \in \mathbb{R}^{T \times C}$. 

The inner operation is a scaled dot-product between these three matrices to encode attention scores as follows:
\begin{equation}
    \mathit{A(Q, K, V)} = \mathrm{Softmax}(\frac{Q.K^T}{\sqrt{P}}).V
\end{equation}

In the above equation, $Q$ and $K$ are replicated matrices from the input to be used inside softmax. A dot-product operation is performed for those matrices as a similarity or correlation measure. Then, softmax is applied to the output of the dot-product, which decides the attention scores from $0$ to $1$. However, these attention scores are discrete and disconnected from the inputs and might cause gradient vanishing. As a result, two more operations are provided to ensure stability in training \cite{transformers} as follows: 1) scaling the output of the dot product by $\sqrt{m}$, which reduces the weight variance. 2) multiplying this scaled-dot product output by matrix $V$ to preserve the spatiality of the input features.

Although this is the main operation of self-attention modules, it is applied through multi-head attention (MHA), which performs simultaneously with various representations. This MHA is achieved by concatenating all the heads:
\begin{equation}
    \mathit{MHA} = \mathit{Concat}(A_i(.)).W, \quad i \in 1, ..., H
\end{equation}
where $W$ is the learned weight of each matrix, whereas $H$ is the number of heads. MHA proves the ability to encode attention from different positions jointly.

The whole self-attention module is then stacked with more layers, including multi-layer perceptron (MLP) and layer normalization \cite{ba2016layer}. At each layer of the transformer $l$, the list of operations is performed in this sequence:
\begin{equation}
    \begin{split}
        &Z_l = \mathit{Mask(MHA(LN(}Z_{l-1}))) + Z_{l-1}, \\
        &Z_l = \mathit{MLP(LN(}Z_l))  + Z_l,\\
        &Z_l = \mathit{LN(}Z_l), \\
        &\mathit{where}\quad l = 1, 2, ..., L 
    \end{split}
    \label{eq:encoder}
\end{equation}
where $LN(.)$ refers to the layer normalization, $l$ is the layer indicator and $L$ is the number of layers on the transformer. In addition, MHA means multi-head attention for the input at layer $l$, whereas $+$ refers to matrix summation. This set of operations can encode all the relationships amongst the input tokens by dividing the input into tokens and then applying these operations to temporally measure the correlations between tokens. Moreover, these operations do not depend on the previous batches of the input, which enables the model to encode the long-range relationships in parallel from multiple batches simultaneously and, as a result, improves the performance significantly.

In this implementation, we mask the self-attention modules to prevent cheating on the following tokens and prevent the attention mechanisms from leaking predictions based on the previous tokens.

\subsection{Classification head}
In this work, we consider binary classification IDSs to classify the input as an attack or not. Transformer encoders are stacked together as shown in Figure \ref{fig:Cst}. The output after $L$ Transformer encoders has the shape $\mathbb{R}^ {B \times T \times C}$ where B is the batch size, T is the token size and C is the channel size. Then, a fully connected layer is used to map it to $\mathbb{R}^{B \times 2}$, where 2 refers to binary classification. 

\subsection{Loss function}
Binary cross entropy is the objective function to penalize the network in the case of wrong predictions as follows:
\begin{equation}
\begin{split}
    CE(y, p) &= -\frac{1}{n} \sum_{i=1}^{n}{y_i}  \log(p_i),
    \end{split}
\label{eq:ce}
\end{equation}
where $y$ refers to the number of ground-truth labels, which is 2 in this study, $p$ denotes the predicted probabilities,

\section{Experiments}
In this part, we discuss the evaluations of PLLM-CS on the network datasets, including UNSW-NB-15 \cite{moustafa2015unsw} and TON\_IoT \cite{alsaedi2020ton_iot}. In the absence of a public dataset for satellite security, these datasets are chosen because they include network data very similar to those of satellites. The settings for the experiments, baselines, datasets and evaluation metrics are provided at the beginning of this section. Then, visual comparisons with state-of-the-art methods are provided to highlight the significance of the PLLM-CS compared to the baselines.

\subsection{Experimental Setup}
PLLM-CS implementation models are implemented on Pytorch, cuDNN, and CUDA-11 on their backends. To guarantee fair comparisons, the same settings are used for all baselines in the training stage.  AdamW is the main optimizer with the same settings as Transformers \cite{transformers}. The server that is used is Quadro GV100 with $32$GB. $30$ is the number of epochs, whereas the learning rate is $2e-5$. Eventually, the remaining settings follow Transformers. The models are used without any fine-tuning or transfer learning.

\subsection{Datasets}
\textbf{UNSW-NB 15:} 
provides network data for anomaly detection in several applications, including satellite security. It contains 49 network features, but we chose 13 of them. The chosen features measure the performance of the network flow, including port, IP, bytes, TTL, load, packets for both source and destination, and duration. It also contains normal data and attack data. Table\ref{tab:unsw} shows the details of this dataset.

\begin{table}[]
\centering
\caption{Description of the network data on UNSW-NB 15}
\begin{tabular}{lll}
\hline
Feature & Type      & Description             \tabularnewline
\hline

srcip   & nominal   & The source IP\tabularnewline
dstip   & nominal   & The destination IP\tabularnewline
proto   & nominal   & The type of protocol\tabularnewline
Sload   & Float     & The source BPS\tabularnewline
Dload   & Float     & The destination BPS\tabularnewline
Stime   & Timestamp & The starting time\tabularnewline
Ltime   & Timestamp & The ending time\tabularnewline
Spkts   & Timestamp & The number of packets\tabularnewline
label   & Binary    & The class of the record\tabularnewline
\hline

\end{tabular}
\label{tab:unsw}
\end{table}

\textbf{TON\_IoT:}
is an IoT dataset that contains telemetry data, network flow data, and operating systems logs. These data were collected for IoT and cybersecurity. It contains some common attacks, including backdoors, Denial of Service (DoS), injection, XSS, scanning, etc. It includes around 22 million records and 45 features. Out of 45 features, we choose eleven, as shown in Table \ref{tab:ton_iot}. 
\begin{table}[]
    \centering
\caption{Description of the network data on TON\_IoT}
\scalebox{1}{
 
\begin{tabular}{lll}
\hline 
Feature  & Type  & Description\tabularnewline
\hline 
time  & time & time of logging data \tabularnewline
date  & date & date of logging data\tabularnewline
motion status  & number & motion sensor status (1 is on and 0 is off)\tabularnewline
light status & boolean & whether the status of light is on or off\tabularnewline
temperature & number & temperature sensor measurement data \tabularnewline
pressure & number & pressure sensor reading \tabularnewline
humidity  & number  & humidity sensor reading\tabularnewline
sphone signal  & boolean  & status of door signal on phone\tabularnewline
latitude  & number  & value of GPS tracker latitude \tabularnewline
longitude  & number  & value of GPS tracker longitude \tabularnewline
FC1 Read Input Register   & number & a code for handling readings of input register \tabularnewline
label  & number  & the class of record as attacked or normal \tabularnewline
\hline 
\end{tabular}
}
    \label{tab:ton_iot}
\end{table}

\begin{table*}[h]
\centering
\caption{Comparisons between PLLM-CS and both types of models, non-trained ML models and trained ML models on UNSW NB-15. The results show that PLLM-CS is outperforming both types of baselines.}
\resizebox{\textwidth}{!}{
\begin{tabular}{l|llllllll}
\hline
&Model & Accuracy & Precision & Recall & F-measure & FNR& AUC & MCC
\tabularnewline
\hline
\multirow{4}{*}{Non-trained ML models}&RF   & 95.2 & 94.8  & 94.9  & 94.9& 6.7 & 94.8& 89.6   \tabularnewline

&ETC   & 95.1 &  94.6& 94.8  &94.7 & 6.8 &94.6 &89.4    \tabularnewline
&XGB & 95.0 & 94.5 &  94.7 & 94.6&7.4  &94.7 &89.0    \tabularnewline
&LGBM & 95.2 &94.6  & 94.9  &94.8 & 7.3 & 94.9& 89.6   \tabularnewline
\hline
\multirow{5}{*}{{Trained ML models}}&CNN   &83.5  &86.9  & 78.1  &80.2 &6.8  &78.2 &64.5    \tabularnewline
&LSTM  &70.6  &67.9  & 66.7       & 67.1&  39.2&66.7 &34.6   \tabularnewline
&BiLSTM &77.2&75.5  &74.5  &74.9   &29.8  & 74.5& 50.0   \tabularnewline
 
&FNN  &65.2  &82.4&51.8  & 42.9    &0&15.6 &51.8    \tabularnewline
&GRU  &74.3  &72.2  & 71.7       & 71.2&  34.7&71.7 &43.9    \tabularnewline
\hline
&\multicolumn{1}{m{1.5cm}}{PLLM-CS}  & 100      & 100       & 100    & 100 &0.0&100&100   \tabularnewline
\hline
\end{tabular}
}
\label{tab:unsw}
\end{table*}

\subsection{Baselines}
The significance of PLLM-CS is evaluated by comparing two cases. The first is traditional classifiers, including the Random Forest (RF), Extra Trees Classifier (ETC), Extreme Gradient Boosting (XGB), and Light Gradient Boosting Method (LGBM), while the second is trained-ML algorithms. The former advanced cyber-security branches during the last decade because of their simplicity and efficiency. However, they are not train-based algorithms, which affects their generalization factor. Also, they are more vulnerable to adversarial attacks than deep learning models. The latter is the preferred way of learning in recent decades, with models trained for a certain number of epochs to learn the patterns of the problem. We compare the PLLM-CS with a CNN, LSTM, BiLSTM, fully-connected NN(FNN) and Gated Recurrent Network (GRU).

\subsection{Evaluation protocols}
We use common evaluation metrics, including recall, precision, F measurement $(F_1)$ and precision, to evaluate the proposed method, PLLM-CS. They are explained in detail as follows:

\textbf{Accuracy:} is one of the basic metrics for evaluating performance. The correct predictions are mainly referred to as $T_p$ in contrast to wrong predictions $T_n$. Accuracy is the ratio of correct samples to the total:

\begin{equation}
    Acc = \frac{T_p + T_n}{T_p + T_n + F_p + F_n}
\end{equation}
\textbf{Precision:}
This metric calculates the score of the true predicted samples $T_p$ compared to the correct ones. 

\begin{equation}
    Prec = \frac{T_p}{T_p +  F_p }
\end{equation}

\textbf{Recall:}
The recall metric is usually used in performance evaluation to calculate how many actual positives the model has predicted. It is important to detect the detection rates of the attacks. 
\begin{equation}
    Rec = \frac{T_p}{T_p +  F_n }
\end{equation}
\textbf{F-measure ($F_1$)}
 is a measure to balance the benefits of Precision and Recall. Compared to Accuracy, $F_1$ is important as a balance between precision and recall, particularly in the cases of uneven distributions for the classes where large numbers of negatives are present.

\begin{equation}
    F-measure = 2 \times \frac{Prec \times Rec}{Prec +  Rec }
\end{equation}

\textbf{False Negative Rate (FNR)} as the name reveals, it is the ratio between false negatives and the total positives (false negatives and true positives)

\begin{equation}
    FNR = \frac{F_n}{F_n + T_p }
\end{equation}
\\
\textbf{Receiver Operating Characteristic (ROC)} is a graphic criterion that plots the true positive rate (TPR) versus the false positive
rate (FPR) with different threshold settings. The chart of ROC uses the following metrics on the x- and y-axis, respectively. 

\begin{align}
    TPR = \frac{T_p}{ T_p + F_n} \\
    FPR = \frac{F_p}{F_p + T_n } 
\end{align}
\\
\textbf{Area Under the ROC Curve (AUC)} is the summation of the performance with all the threshold settings. It is a desirable measure because it is scale-invariant and classification-threshold invariant, which measures how well the predictions are ranked and how irrelevant they are to the classification's threshold.
\\
\textbf{Matthews Correlation Coefficient (MCC)} is considered the best classification metric that summarizes the confusion matrix and considers the four metrics TP, TF, FP, and FN.

\begin{equation}
    MCC = \frac{T_p \times T_n - F_p \times F_n}{\sqrt{(T_p+F_p)(T_p+F_n)(T_n + F_p)(T_n + F_n)} }
\end{equation}

In the above settings, $T_p$ refers to true positives, $T_n$ represents true negatives, false positives are $F_p$, whereas false negatives are $F_n$.

\textbf{Baselines settings}
For UNSW-NB 15, we chose non-trained models such as Random Forrest, ETC, XGB, and LGBM. Also, deep learning models include convolution neural network (CNN), long-short-term model (LSTM), Bidirectional LSTM (BiLSTM), fully connected neural network (FNN), and gated recurrent unit (GRU). For all the models, the number of epochs is $100$, the batch size is $16$, the feature size is $32$, and the learning rate is $1 \times 10^{-3}$. The FNN model has three main linear layers and two non-linear layers (ReLU \cite{agarap2018deep}), while the last is the classifier. CNN model has two main layers, each composed of 1D convolution, batch normalization \cite{pmlr-v37-ioffe15}, and ReLU, whereas the last layer is the classifier. The feature map size is $64$ and $128$, respectively. Recurrent models include two layers with a size of $64$ for LSTM and GRU, whereas BiLSTM is $128$. For the proposed model, the number of epochs is $10$, and the number of Transformers blocks is $2$. The dimension of the attention block is $32$ with $4$ heads. The loss function is cross entropy for all the variants, whereas the optimizer is AdamW \cite{loshchilov2018decoupled}.

\begin{figure}[]
  \begin{subfigure}[]{.48\columnwidth}
    \centering\includegraphics[width=\linewidth]{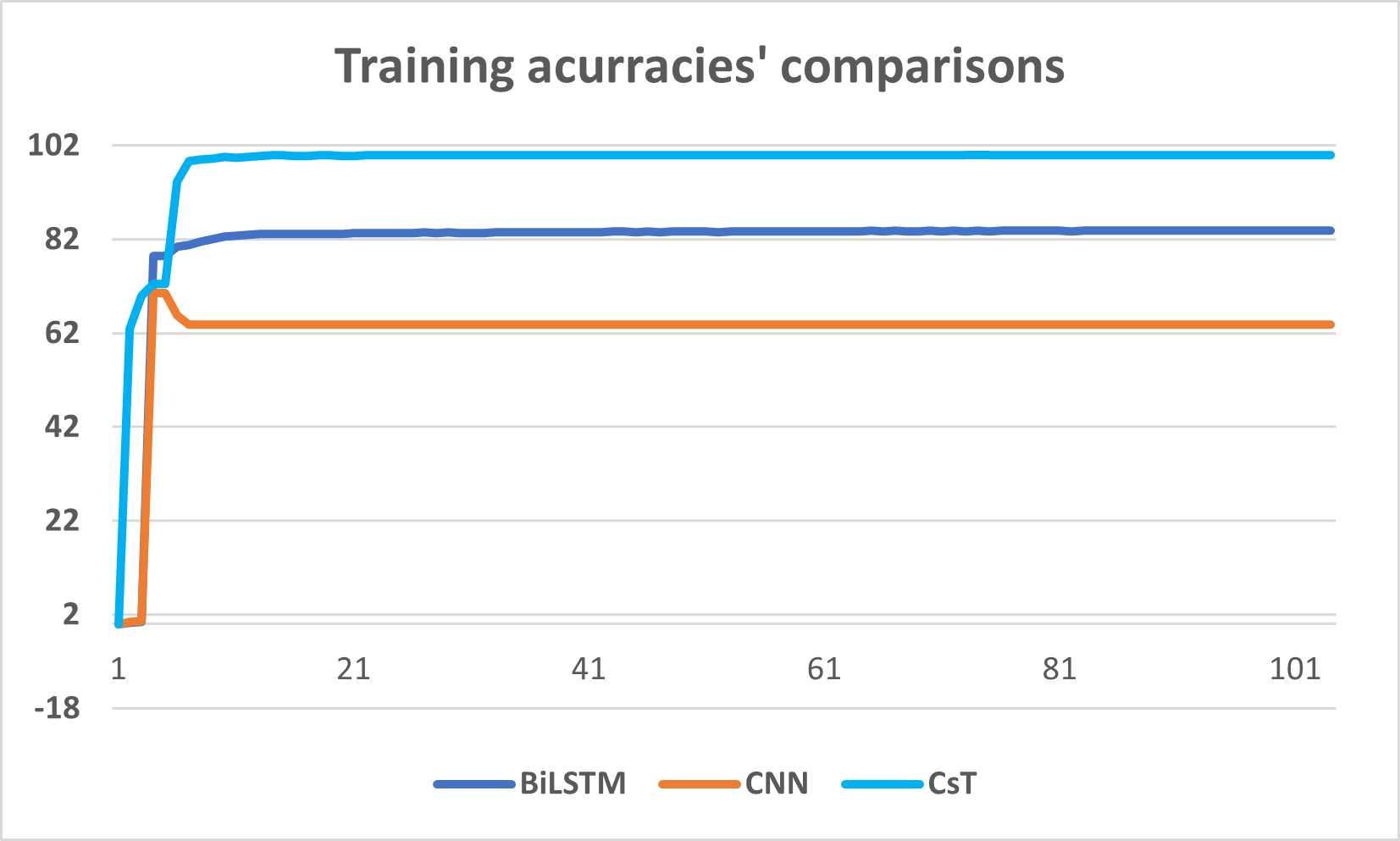}
    \caption{Training}
  \end{subfigure}
  \begin{subfigure}[]{.48\columnwidth}
    \centering\includegraphics[width=\linewidth]{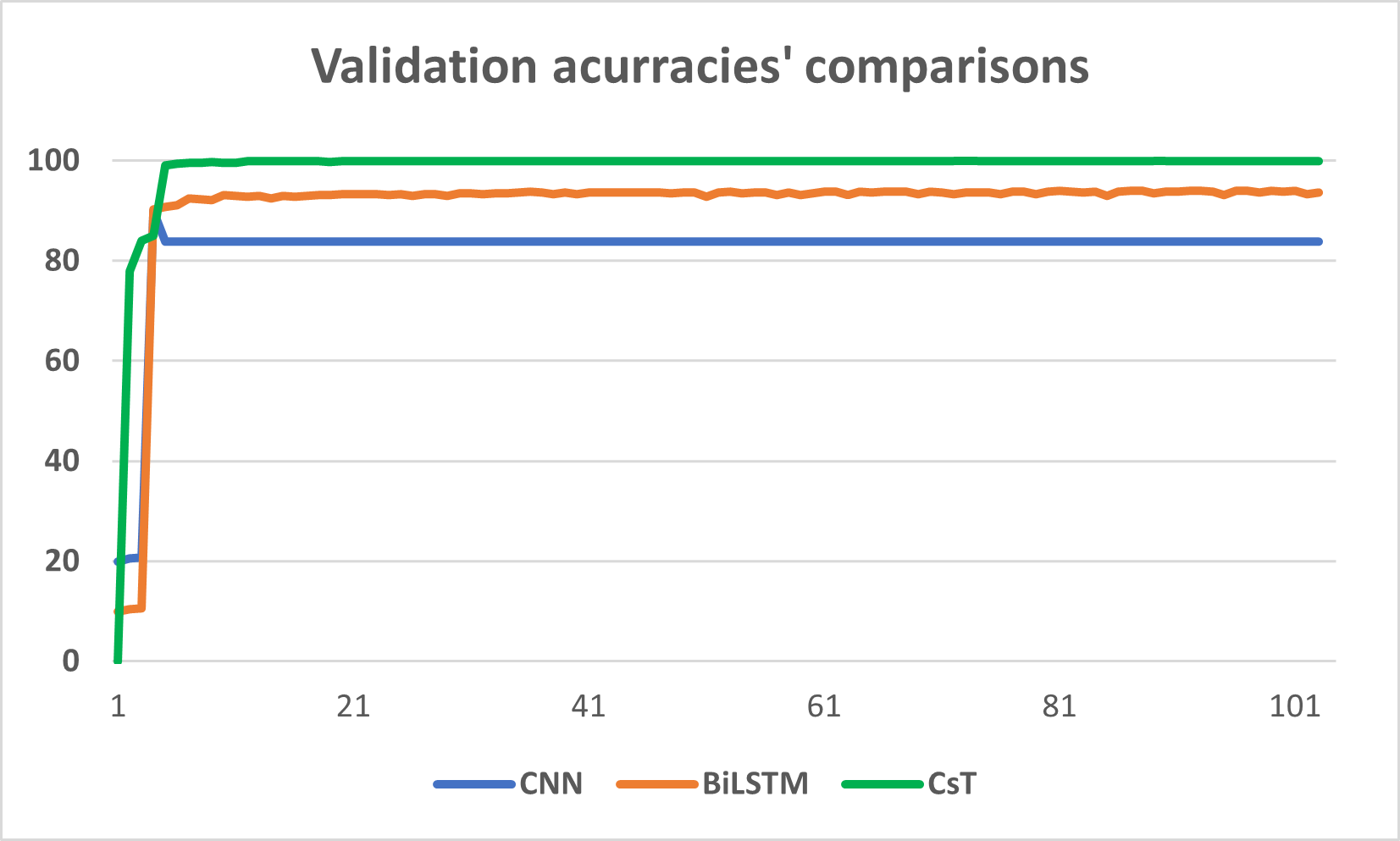}
   \caption{Validation}
   \end{subfigure}
  \caption{Visual comparisons between PLLM-CS, CNN, and BiLSTM on the training (left) and testing (right) phases on the UNSW NB-15 dataset. The proposed method's accuracy is higher than the baselines.}
    \label{fig:accuracy_unsw}
\end{figure}
\textbf{UNSW NB-15 Results}
In this part, we discuss the experimental results of the UNSW NB-15 dataset between the proposed method, PLLM-CS, and the baselines (RF, ETC, XGB, LGBM, CNN, LSTM, BiLSTM, FNN, and GRU). The results are illustrated in Table \ref{tab:unsw}. 
Firstly, comparing PLLM-CS to non-trained machine learning models \ie traditional machine learning classifiers, the proposed method obtains better accuracy from the first few training epochs. Though these methods are simple in architecture and complexity, they provide high accuracy, up to $95.2\%$. The improvement amongst such models is very minor. For instance, the difference between RF and ETC is only $0.1\%$. Overall, PLLM-CS obtains better than all non-trained models with a large margin of $4.8\%$. Compared to deep learning models \eg CNN, LSTM, BiLSTM, GRU, and FNN, PLLM-CS shows great significance. It obtains the highest accuracy $100\%$ in all the metrics, while the second best is LSTM, with 93.0 accuracies. Though deep learning models are inferior to pre-deep learning models, trained models are generalizing better than non-trained ones. Overall, the proposed method outperforms both types with a considerable margin of $4.9\%$ compared to the closest one. Diving into details, the metrics of the comparisons are accuracy, precision, recall, f-measure, AUC, and MCC. The proposed method showed significance in all the metrics. For example, in accuracy, it is the highest at $99.9$. Also, F-measure, recall, precision, and MCC reported better results with PLLM-CS. It is worth noting that LSTM behaves better than CNN and FNN because of its ability to encode the context better than the input tokens.
The reason behind this is encoding the context in the feature space. This is because of the presence of self-attention modules that are capable of learning long-range relationships.

\begin{figure}[]
  \begin{subfigure}[]{.48\columnwidth}
    \centering\includegraphics[width=\linewidth]{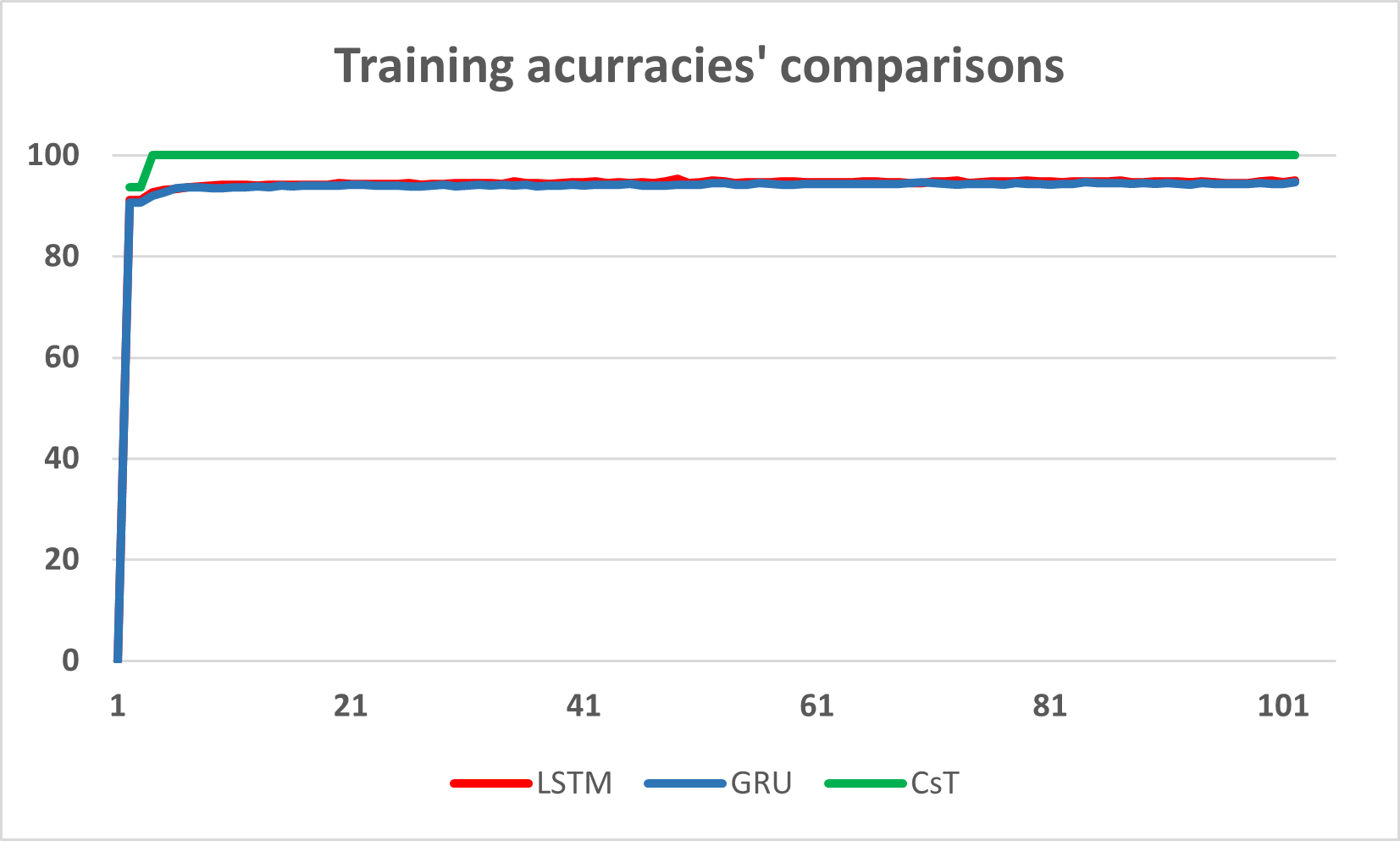}
    \caption{Training}
  \end{subfigure}
  \begin{subfigure}[]{.48\columnwidth}
    \centering\includegraphics[width=\linewidth]{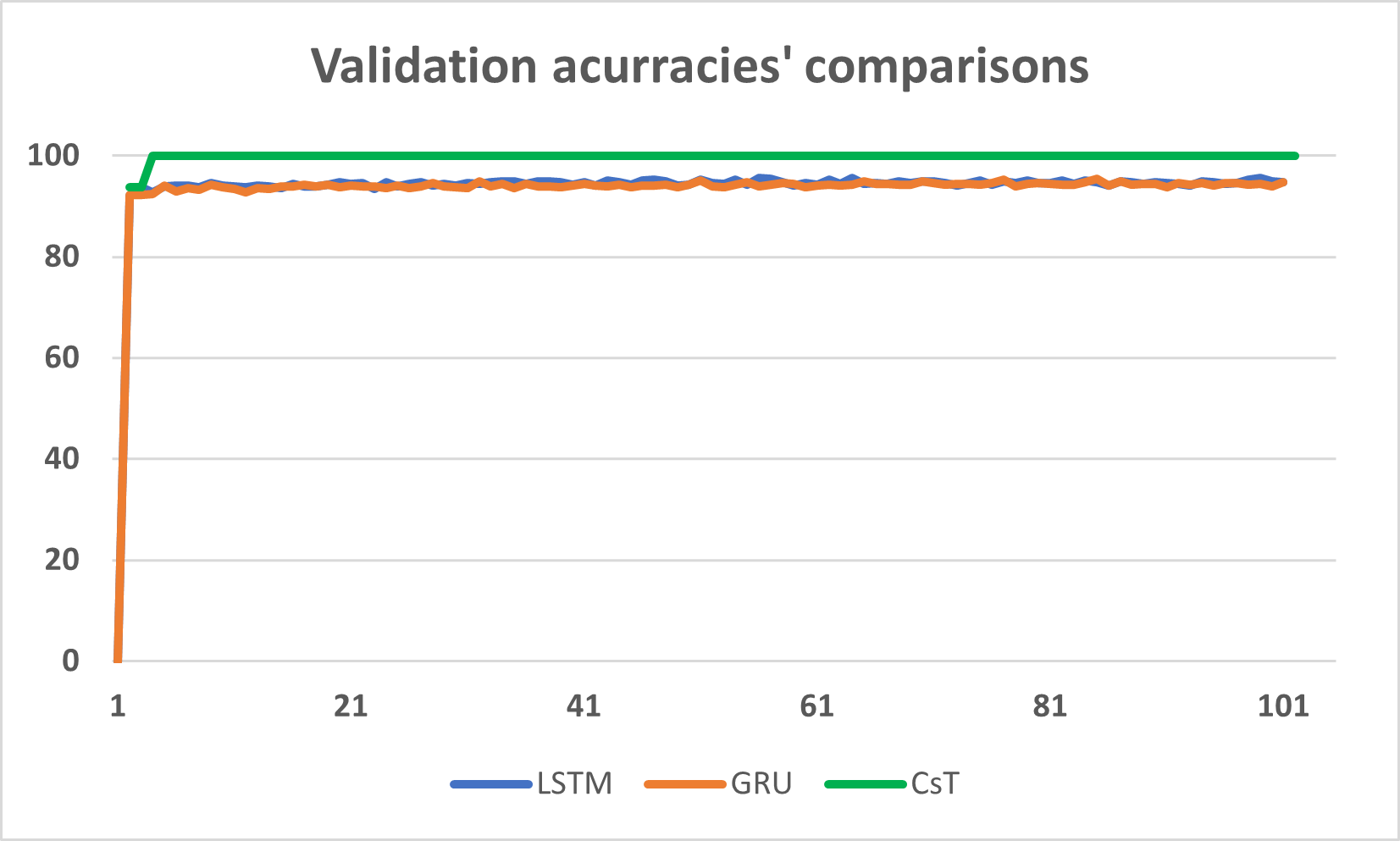}
   \caption{Validation}
   \end{subfigure}
  \caption{Visual comparisons between PLLM-CS, GRU, and LSTM on the training (left) and testing (right) phases on the TON\_IoT dataset. The proposed method's accuracy is higher than the baselines.}
    \label{fig:accuracy_ton}
\end{figure}

\begin{figure}[]
  \begin{subfigure}[]{.48\columnwidth}
    \centering\includegraphics[width=\linewidth]{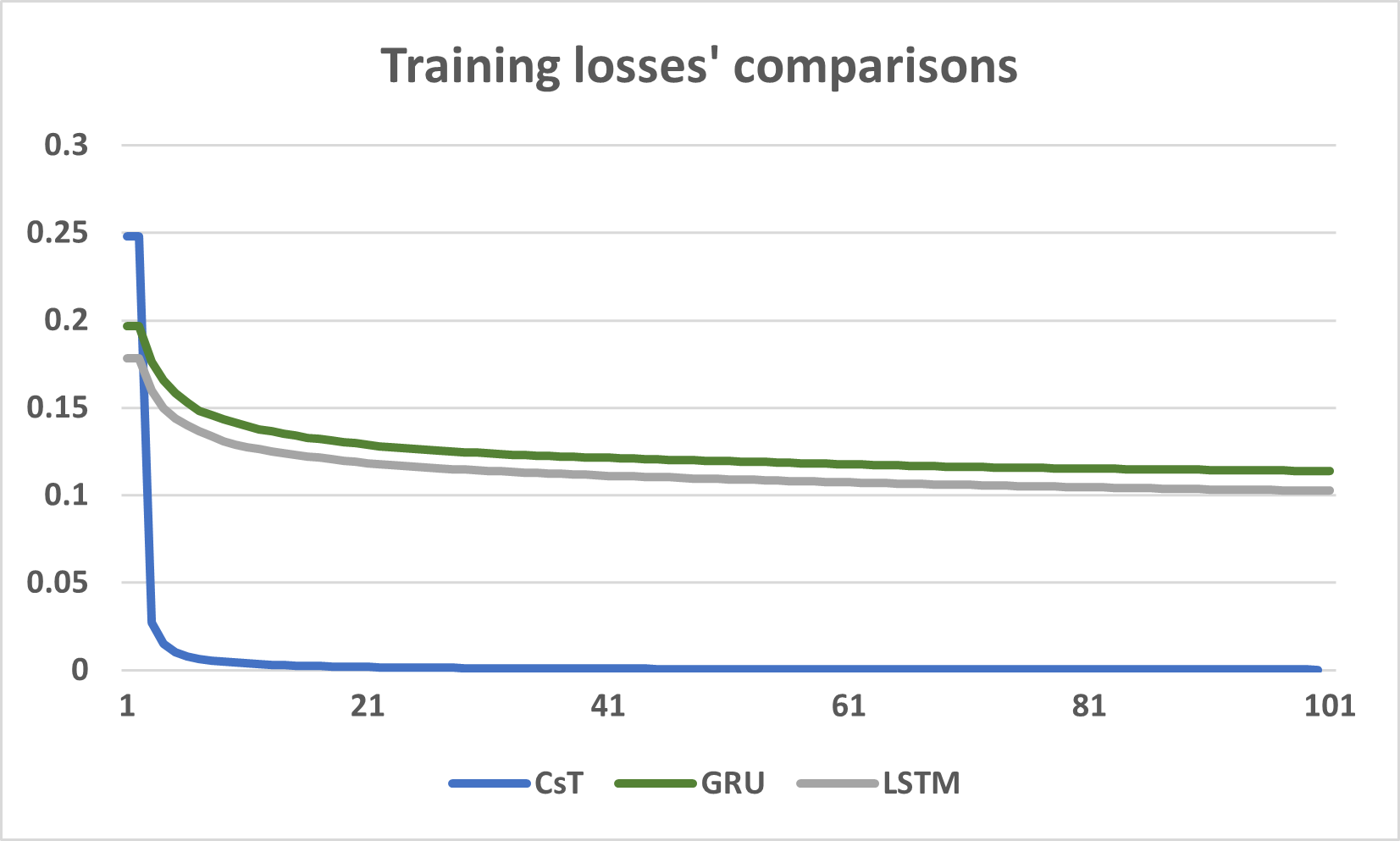}
    \caption{Training}
  \end{subfigure}
  \begin{subfigure}[]{.48\columnwidth}
    \centering\includegraphics[width=\linewidth]{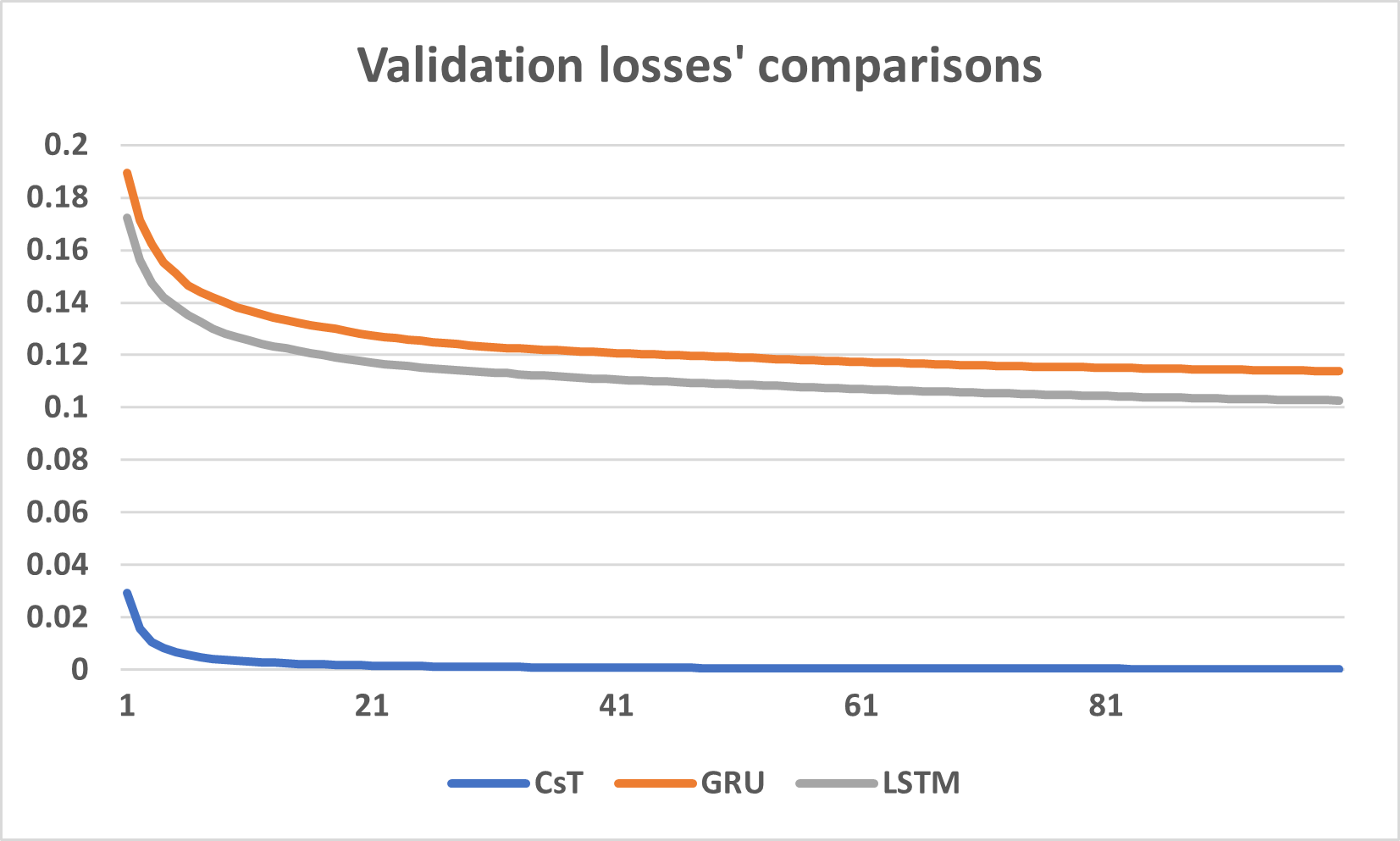}
   \caption{Validation}
   \end{subfigure}
  \caption{Visual comparisons of losses for PLLM-CS, CNN, and BiLSTM on the training (left) and testing (right) phases on the TON\_IoT dataset. The proposed method shows stable convergence.}
  \label{fig:losses_ton}
\end{figure}

The visual comparisons between the proposed methods and the previous deep learning models (\eg CNN and BiLSTM) on TON\_IoT are shown in Figures \ref{fig:accuracy_ton} and \ref{fig:losses_ton}. PLLM-CS shows significant validation accuracy with a large margin during the stability of the training. More precisely, Figure \ref{fig:losses_ton} shows that the behaviour of PLLM-CS in loss convergence is far better than the baselines. PLLM-CS is achieving higher performance than CNN and LSTM with better loss convergence.

\begin{table*}[]
\centering
\caption{Comparisons between PLLM-CS and both types of models, non-trained ML models and trained ML models on TON\_IoT. The results show that PLLM-CS is outperforming both types of baselines.}
\resizebox{\textwidth}{!}{
\begin{tabular}{l|l|lllllll}
\hline
&Model & Accuracy & Precision & Recall & F-measure & FNR& AUC & MCC
\tabularnewline
\hline
\multirow{4}{*}{Non-trained ML models}&RF   & 100.0 &  100.0& 100.0  &100.0 & 0 &100.0 &100.0    \tabularnewline

&ETC   & 100.0 &  100.0& 100.0  &100.0 & 0 &100.0 &100.0    \tabularnewline
&XGB & 100.0 &  100.0& 100.0  &100.0 & 0 &100.0 &100.0    \tabularnewline
&LGBM & 100.0 &  100.0& 100.0  &100.0 & 0 &100.0 &100.0    \tabularnewline
\hline
\multirow{5}{*}{Trained ML models}&CNN   & 98.27  &98.35  & 97.9  &98.12 &1.8  & 97.9&96.8    \tabularnewline

&LSTM &94.6&94.7  &94.1  & 94.1   &4.1  &94.1 &88.2    \tabularnewline
&BiLSTM  & 94.5  &93.9  & 94.5       & 94.2&3.3  & 94.5&  88.5  \tabularnewline 
&FNN  &90.78  &93.37&86.88  & 89.05    &12.2&86.88 &79.99    \tabularnewline
&GRU  &94.7 &93.5&95.2&94.2 &1.5& 95.3&88.8    \tabularnewline
\hline
&\multicolumn{1}{m{2cm}}{PLLM-CS}  & 100.0      & 100.0       & 100.0    & 100.0 &0&100.0 &100.0   \tabularnewline
\hline

\end{tabular}
}
\label{tab:ton}
\end{table*}

\begin{figure}[]
  \begin{subfigure}[]{.48\columnwidth}
    \centering\includegraphics[width=\linewidth]{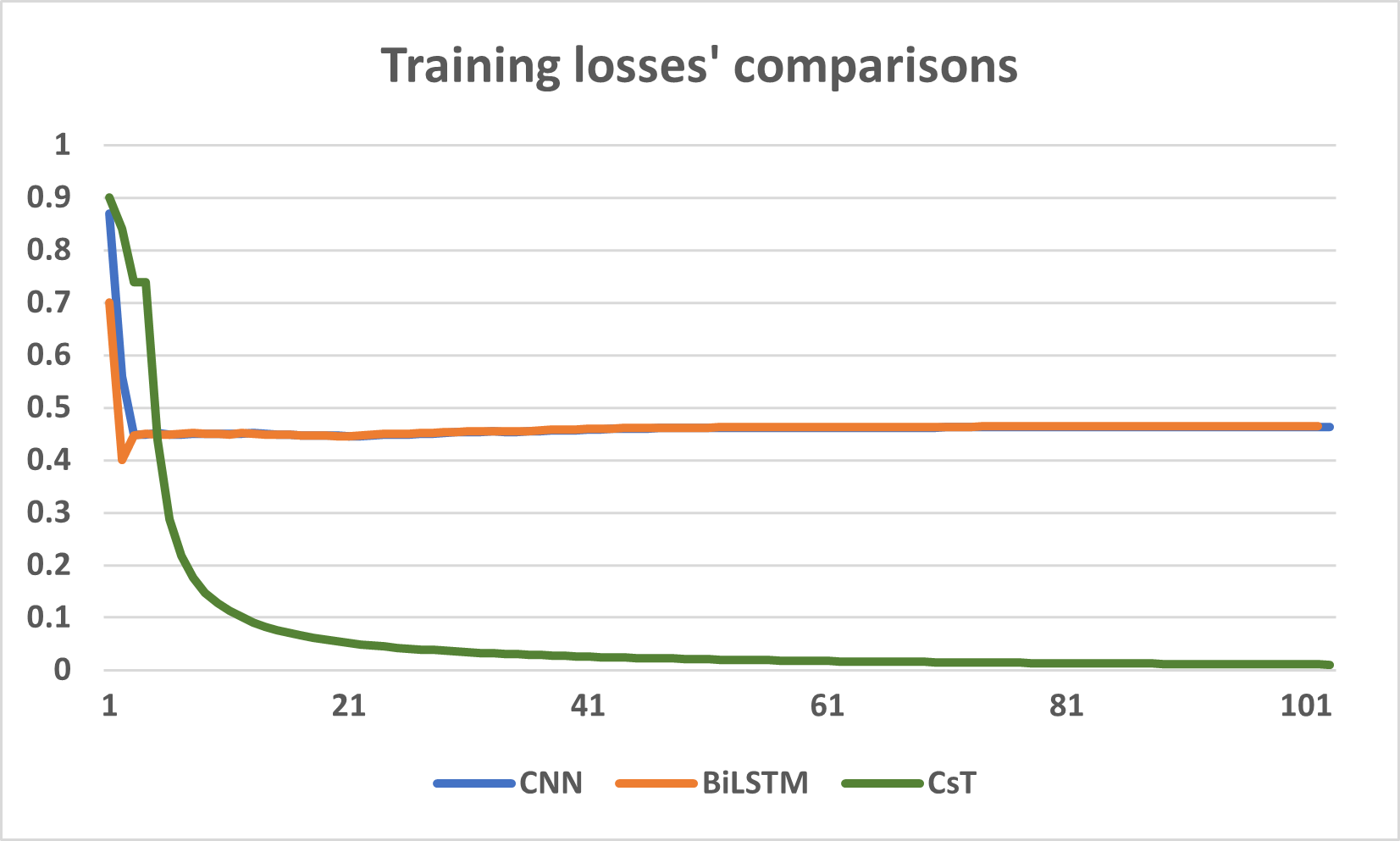}
    \caption{Training}
  \end{subfigure}
  \begin{subfigure}[]{.48\columnwidth}
    \centering\includegraphics[width=\linewidth]{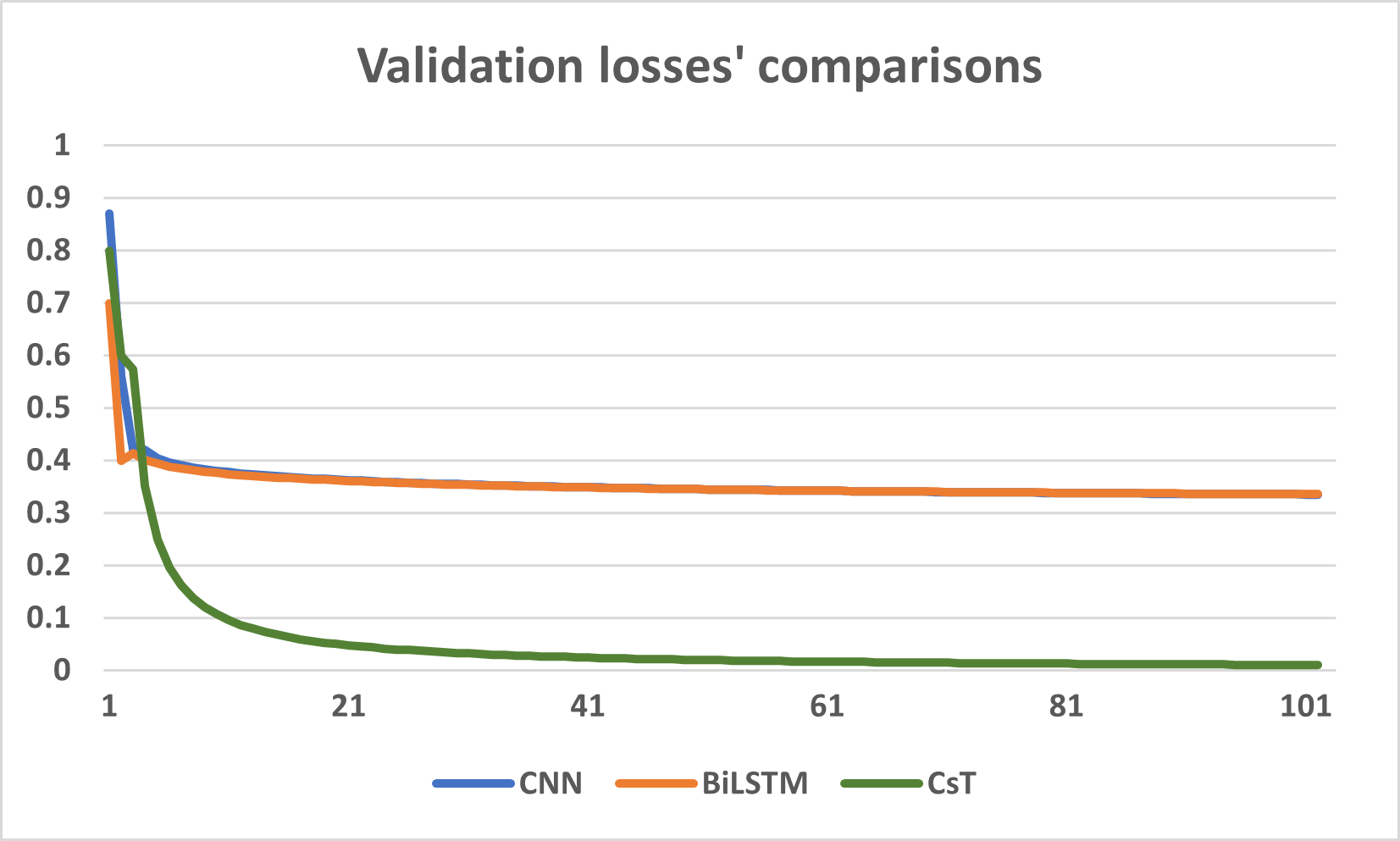}
   \caption{Validation}
   \end{subfigure}
  \caption{Visual comparisons of losses for PLLM-CS, CNN, and BiLSTM on the training (left) and testing (right) phases on the UNSW NB-15 dataset. The proposed method shows stable convergence.}
  \label{fig:losses_unsw}
\end{figure}

\textbf{TON IOT Results:}
In this part, we discuss the experimental results on the TON\_IoT dataset between the proposed method, PLLM-CS, and the baselines (RF, ETC, XGB, LGBM, CNN, LSTM, BiLSTM, FNN, and GRU). In Table \ref{tab:ton}, the results are illustrated. 
Firstly, PLLM-CS was compared to non-trained machine learning models. The proposed method obtained the highest accuracy, 100\%, after a few training epochs. Though these methods are simple in architecture and complexity, they provide high accuracy, $100\%$. These models achieved a $0$ False Negative Rate (FNR).  Overall, PLLM-CS behaves better than non-trained models even though they achieved the full mark. This is because PLLM-CS is a trained-based algorithm that behaves better with novel examples and can generalise. Compared to deep learning models \eg CNN, LSTM, BiLSTM, GRU, and FNN, PLLM-CS shows great significance. It obtains the highest accuracy $100\%$ in all the metrics, while the second best is CNN with 98.27 accuracies. Though deep learning models are inferior to pre-deep learning models, trained models are generalizing better than non-trained ones. Overall, the proposed method outperforms both types. Diving into details, the metrics of the comparisons are accuracy, precision, recall, f-measure, AUC, and MCC. The proposed method showed significance in all the metrics. For example, in accuracy, it is the highest with $100\%$. Also, F-measure, recall, precision, and MCC reported better results with PLLM-CS. 
Again, it is proved that PLLM-CS behaves better because it can encode the context in the feature space.

The visual comparisons between the proposed methods and the previous deep learning models (\eg CNN and LSTM) on TON\_IoT are shown in Figures \ref{fig:accuracy_unsw} and \ref{fig:losses_unsw}. PLLM-CS shows significant validation accuracy with a large margin while the training stability. More precisely, Figure \ref{fig:losses_unsw} shows that the behaviour of PLLM-CS in loss convergence is far better than the baselines. PLLM-CS is achieving higher performance than CNN and LSTM with better loss convergence.

Conclusively, PLLM-CS provides a great advantage in providing robustness for SSNs. This study provides Transformer-based IDSs for SSNs. 
As shown in the experiments section, PLLM-CS illustrates significant accuracy in two publicly available datasets, which include network data from diverse IoT domains. Also, PLLM-CS performance achieved the full mark of $100\%$ without any fine-tuning or hyper-parameter optimization.

\section{Conclusion and future work}
In this study, a simple yet efficient intrusion detection model using contextual transformers, the PLLM-CS, to detect intrusions on the network data of SSNs is proposed. It adapts transformers to suit cyber-security datasets by sentencing the input data, enabling them to encode long-term relationships. This is the first study to use transformers and attention-based models to detect intrusions on SSNs. The empirical results obtained from two real datasets, UNSW-NB 15 and TON\_IoT, show the superiority of the proposed model over the baselines of RF, XGB, CNN, FNN, GRU, LSTM and BiLSTM. We conclude that this is because of its capability to encode contextual information using self-attention modules. However, as a satellite system does not have a proper dataset that mimics real data, the PLLM-CS’s real-time network data used for testing are similar to satellite ones. Therefore, a future direction would be to develop a special dataset for SSNs and another to consider the constraint of the limited power inside them that requires efficient algorithms to increase their speeds and power consumption.

 \bibliographystyle{elsarticle-num} 
 \bibliography{LLM.bib}

\begin{thebibliography}{10}
\expandafter\ifx\csname url\endcsname\relax
  \def\url#1{\texttt{#1}}\fi
\expandafter\ifx\csname urlprefix\endcsname\relax\def\urlprefix{URL }\fi
\expandafter\ifx\csname href\endcsname\relax
  \def\href#1#2{#2} \def\path#1{#1}\fi

\bibitem{transformers}
A.~Vaswani, N.~Shazeer, N.~Parmar, J.~Uszkoreit, L.~Jones, A.~N. Gomez, {\L}.~Kaiser, I.~Polosukhin, Attention is all you need, Advances in neural information processing systems 30 (2017).

\bibitem{de2015satellite}
M.~De~Sanctis, E.~Cianca, G.~Araniti, I.~Bisio, R.~Prasad, Satellite communications supporting internet of remote things, IEEE Internet of Things Journal 3~(1) (2015) 113--123.

\bibitem{badue2021self}
C.~Badue, R.~Guidolini, R.~V. Carneiro, P.~Azevedo, V.~B. Cardoso, A.~Forechi, L.~Jesus, R.~Berriel, T.~M. Paixao, F.~Mutz, et~al., Self-driving cars: A survey, Expert Systems with Applications 165 (2021) 113816.

\bibitem{colagrossi2022spacecraft}
A.~Colagrossi, M.~Lavagna, A spacecraft attitude determination and control algorithm for solar arrays pointing leveraging sun angle and angular rates measurements, Algorithms 15~(2) (2022) 29.

\bibitem{iot_risks}
\href{https://start.paloaltonetworks.com/unit-42-iot-threat-report}{Iot adoption and its security risks have both grown}.
\newline\urlprefix\url{https://start.paloaltonetworks.com/unit-42-iot-threat-report}

\bibitem{satellite_risks}
\href{https://www.aljazeera.com/news/2022/5/10/russia-behind-cyberattack-against-internet-network-in-ukraine}{Russia downed satellite internet in ukraine: Western officials}.
\newline\urlprefix\url{https://www.aljazeera.com/news/2022/5/10/russia-behind-cyberattack-against-internet-network-in-ukraine}

\bibitem{kolias2017ddos}
C.~Kolias, G.~Kambourakis, A.~Stavrou, J.~Voas, Ddos in the iot: Mirai and other botnets, Computer 50~(7) (2017) 80--84.

\bibitem{wankhede2018attack}
S.~Wankhede, D.~Kshirsagar, Dos attack detection using machine learning and neural network, in: ICCUBEA, IEEE, 2018, pp. 1--5.

\bibitem{verisign}
{Stats of cars}, \url{https://www.verisign.com/enIN/securityservices/ddosprotection/ddosreport/index.xhtml}, [Online; accessed 27-April-2022] (2022).

\bibitem{lau2000distributed}
F.~Lau, S.~H. Rubin, M.~H. Smith, L.~Trajkovic, Distributed denial of service attacks, Vol.~3, IEEE, 2000, pp. 2275--2280.

\bibitem{wood2002denial}
A.~D. Wood, J.~A. Stankovic, Denial of service in sensor networks, computer 35~(10) (2002) 54--62.

\bibitem{borisov2007denial}
N.~Borisov, G.~Danezis, P.~Mittal, P.~Tabriz, Denial of service or denial of security?, in: Proceedings of the 14th ACM conference on Computer and communications security, 2007, pp. 92--102.

\bibitem{li2020distributed}
K.~Li, H.~Zhou, Z.~Tu, W.~Wang, H.~Zhang, Distributed network intrusion detection system in satellite-terrestrial integrated networks using federated learning, IEEE Access 8 (2020) 214852--214865.

\bibitem{moustafa2022dfsat}
N.~Moustafa, I.~A. Khan, M.~Hassanin, D.~Ormrod, D.~Pi, I.~Razzak, J.~Slay, Dfsat: Deep federated learning for identifying cyber threats in iot-based satellite networks, IEEE Transactions on Industrial Informatics (2022).

\bibitem{jackson2018exploring}
S.~Jackson, J.~Straub, S.~Kerlin, Exploring a novel cryptographic solution for securing small satellite communications., Int. J. Netw. Secur. 20~(5) (2018) 988--997.

\bibitem{o2016secure}
M.~O'Neill, E.~O'Sullivan, G.~McWilliams, M.-J. Saarinen, C.~Moore, A.~Khalid, J.~Howe, R.~Del~Pino, M.~Abdalla, F.~Regazzoni, et~al., Secure architectures of future emerging cryptography safecrypto, in: Proceedings of the ACM International Conference on Computing Frontiers, 2016, pp. 315--322.

\bibitem{ostad2019efficient}
A.~Ostad-Sharif, D.~Abbasinezhad-Mood, M.~Nikooghadam, Efficient utilization of elliptic curve cryptography in design of a three-factor authentication protocol for satellite communications, Computer Communications 147 (2019) 85--97.

\bibitem{zhao2020intelligent}
R.~Zhao, Y.~Yin, Y.~Shi, Z.~Xue, Intelligent intrusion detection based on federated learning aided long short-term memory, Physical Communication 42 (2020) 101157.

\bibitem{studer2009coremelt}
A.~Studer, A.~Perrig, The coremelt attack, in: European Symposium on Research in Computer Security, Springer, 2009, pp. 37--52.

\bibitem{kang2013crossfire}
M.~S. Kang, S.~B. Lee, V.~D. Gligor, The crossfire attack, in: 2013 IEEE symposium on security and privacy, IEEE, 2013, pp. 127--141.

\bibitem{giuliari2021icarus}
G.~Giuliari, T.~Ciussani, A.~Perrig, A.~Singla, $\{$ICARUS$\}$: Attacking low earth orbit satellite networks, in: 2021 USENIX Annual Technical Conference (USENIX ATC 21), 2021, pp. 317--331.

\bibitem{na2018distributed}
Z.~Na, Z.~Pan, X.~Liu, Z.~Deng, Z.~Gao, Q.~Guo, Distributed routing strategy based on machine learning for leo satellite network, Wireless Communications and Mobile Computing 2018 (2018).

\bibitem{gunn2018anomaly}
L.~Gunn, P.~Smet, E.~Arbon, M.~D. McDonnell, Anomaly detection in satellite communications systems using lstm networks, in: 2018 Military Communications and Information Systems Conference (MilCIS), IEEE, 2018, pp. 1--6.

\bibitem{pilastre2020anomaly}
B.~Pilastre, L.~Boussouf, S.~d’Escrivan, J.-Y. Tourneret, Anomaly detection in mixed telemetry data using a sparse representation and dictionary learning, Signal Processing 168 (2020) 107320.

\bibitem{cheng2021research}
F.~Cheng, X.~Guo, Y.~Qi, J.~Xu, W.~Qiu, Z.~Zhang, W.~Zhang, N.~Qi, Research on satellite power anomaly detection method based on lstm, in: 2021 IEEE International Conference on Power Electronics, Computer Applications (ICPECA), IEEE, 2021, pp. 706--710.

\bibitem{wang2022deep}
Y.~Wang, J.~Gong, J.~Zhang, X.~Han, A deep learning anomaly detection framework for satellite telemetry with fake anomalies, International Journal of Aerospace Engineering 2022 (2022).

\bibitem{zeng2022satellite}
Z.~Zeng, G.~Jin, C.~Xu, S.~Chen, Z.~Zeng, L.~Zhang, Satellite telemetry data anomaly detection using causal network and feature-attention-based lstm, IEEE Transactions on Instrumentation and Measurement 71 (2022) 1--21.

\bibitem{yun2022data}
S.-T. Yun, S.-H. Kong, Data-driven in-orbit current and voltage prediction using bi-lstm for leo satellite lithium-ion battery soc estimation, IEEE Transactions on Aerospace and Electronic Systems (2022).

\bibitem{hassanin2022crossformer}
M.~Hassanin, A.~Khamiss, M.~Bennamoun, F.~Boussaid, I.~Radwan, Crossformer: Cross spatio-temporal transformer for 3d human pose estimation, arXiv preprint arXiv:2203.13387 (2022).

\bibitem{hassanin2022visual}
M.~Hassanin, S.~Anwar, I.~Radwan, F.~S. Khan, A.~Mian, Visual attention methods in deep learning: An in-depth survey, arXiv preprint arXiv:2204.07756 (2022).

\bibitem{tan2019neural}
M.~Tan, A.~Iacovazzi, N.-M.~M. Cheung, Y.~Elovici, A neural attention model for real-time network intrusion detection, in: LCN, IEEE, 2019, pp. 291--299.

\bibitem{wu2022rtids}
Z.~Wu, H.~Zhang, P.~Wang, Z.~Sun, Rtids: a robust transformer-based approach for intrusion detection system, IEEE Access (2022).

\bibitem{ghourabi2022security}
A.~Ghourabi, A security model based on lightgbm and transformer to protect healthcare systems from cyberattacks, IEEE Access 10 (2022) 48890--48903.

\bibitem{luo2022hierarchical}
S.~Luo, Z.~Zhao, Q.~Hu, Y.~Liu, A hierarchical cnn-transformer model for network intrusion detection, in: CAMMIC, Vol. 12259, SPIE, 2022, pp. 853--860.

\bibitem{zaheer2020big}
M.~Zaheer, G.~Guruganesh, K.~A. Dubey, J.~Ainslie, C.~Alberti, S.~Ontanon, P.~Pham, A.~Ravula, Q.~Wang, L.~Yang, et~al., Big bird: Transformers for longer sequences, Advances in Neural Information Processing Systems 33 (2020) 17283--17297.

\bibitem{hassanin2019new}
M.~Hassanin, S.~Khan, M.~Tahtali, A new localization objective for accurate fine-grained affordance segmentation under high-scale variations, IEEE Access 8 (2019) 28123--28132.

\bibitem{ba2016layer}
J.~L. Ba, J.~R. Kiros, G.~E. Hinton, Layer normalization, arXiv preprint arXiv:1607.06450 (2016).

\bibitem{moustafa2015unsw}
N.~Moustafa, J.~Slay, Unsw-nb15: a comprehensive data set for network intrusion detection systems (unsw-nb15 network data set), in: MilCIS, IEEE, 2015, pp. 1--6.

\bibitem{alsaedi2020ton_iot}
A.~Alsaedi, N.~Moustafa, Z.~Tari, A.~Mahmood, A.~Anwar, Ton\_iot telemetry dataset: A new generation dataset of iot and iiot for data-driven intrusion detection systems, IEEE Access 8 (2020) 165130--165150.

\bibitem{agarap2018deep}
A.~F. Agarap, Deep learning using rectified linear units (relu), arXiv preprint arXiv:1803.08375 (2018).

\bibitem{pmlr-v37-ioffe15}
S.~Ioffe, C.~Szegedy, Batch normalization: Accelerating deep network training by reducing internal covariate shift, in: F.~Bach, D.~Blei (Eds.), Proceedings of the 32nd International Conference on Machine Learning, Vol.~37 of Proceedings of Machine Learning Research, PMLR, Lille, France, 2015, pp. 448--456.

\bibitem{loshchilov2018decoupled}
I.~Loshchilov, F.~Hutter, Decoupled weight decay regularization, in: International Conference on Learning Representations, 2018.

\end{thebibliography}





\end{document}